\documentclass[runningheads]{llncs}
\usepackage[T1]{fontenc}
\usepackage{amsmath}
\usepackage{graphicx}
\begin{document}
\title{Advancing Financial Forecasting: A Comparative Analysis of Neural Forecasting Models N-HiTS and N-BEATS}

\titlerunning{Neural Forecasting Models for Financial Markets}
%
\author{Mohit Apte\inst{1}\orcidID{0009-0005-4249-3151} \and
Dr. Y.V. Haribhakta\inst{2}\orcidID{0000-0003-3882-4091}}
\authorrunning{M. Apte et al.}
%
\institute{Computer Engineering, COEP Technological University, Pune, India \\
\email{aptemp21.comp@coeptech.ac.in} \and
Asst. Prof, Dept. of Computer Engineering, COEP Technological University, Pune, India \\
\email{ybl.comp@coeptech.ac.in}}
\maketitle              

\begin{abstract}
In the rapidly evolving field of financial forecasting, the application of neural networks presents a compelling advancement over traditional statistical models. This research paper explores the effectiveness of two specific neural forecasting models, N-HiTS and N-BEATS, in predicting financial market trends. Through a systematic comparison with conventional models, this study demonstrates the superior predictive capabilities of neural approaches, particularly in handling the non-linear dynamics and complex patterns inherent in financial time series data. The results indicate that N-HiTS and N-BEATS not only enhance the accuracy of forecasts but also boost the robustness and adaptability of financial predictions, offering substantial advantages in environments that require real-time decision-making. The paper concludes with insights into the practical implications of neural forecasting in financial markets and recommendations for future research directions.
\end{abstract}

	\section{Introduction}
Financial markets are intricate systems shaped by a wide array of elements, from macroeconomic indicators to detailed trading activities at the micro level. Traditional statistical models, while useful, often fall short in capturing the non-linear and intricate dependencies within market data. This limitation prompts the exploration of more advanced techniques capable of handling such complexities. Neural forecasting models, particularly those designed for time series data, offer promising alternatives due to their ability to learn and model the underlying patterns of financial markets dynamically.

The introduction of neural networks in forecasting has revolutionized many areas of financial analytics, providing tools that can adapt to and learn from the data in a way that static models cannot. Among these tools, the N-HiTS and N-BEATS models stand out due to their architectural innovations and proven effectiveness in various forecasting competitions and applications. These models harness deep learning to deliver forecasts that are not only accurate but also capable of dealing with the volatility and variability inherent in financial time series.

This paper aims to:
\begin{itemize}
    \item Evaluate the performance of N-HiTS and N-BEATS in the context of financial market forecasting, comparing their effectiveness against traditional statistical models.

    \item Demonstrate how these neural models can be utilized to enhance the accuracy and reliability of financial predictions.

    \item Provide empirical evidence and theoretical insights that substantiate the superiority of neural approaches in complex forecasting environments
\end{itemize}

\section{Literature Review}
The pursuit of accurate financial forecasting has historically leveraged statistical methods, including ARIMA and exponential smoothing. These methods have proven effective under conditions of linear relationships and minimal structural breaks in data. However, the financial markets are characterized by their non-linear interactions and abrupt changes, challenging the adequacy of traditional models.

\subsection{Statistical Models in Financial Forecasting:}
\begin{itemize}
    \item ARIMA (Autoregressive Integrated Moving Average): Widely used for time series prediction, ARIMA models financial time series by explaining the autocorrelations in data. Despite its popularity, ARIMA struggles with non-linear patterns and market shocks (Box \& Jenkins, 1976)\cite{b1}.
    \item Exponential Smoothing: This method, including its variants like Holt-Winters, provides a framework to forecast data with trends and seasonality, adjusting to changes with smoothing constants. However, its simplicity can be a drawback in complex financial scenarios (Holt, 1957; Winters, 1960)\cite{b2,b3}.

\end{itemize}

\subsection{Rise of Machine Learning in Forecasting:}
Recent decades have seen a shift towards machine learning (ML) techniques that offer robustness against the market’s stochastic nature. ML models, particularly those employing neural networks, are celebrated for their capacity to model non-linearities and interactions without explicitly predefined equations.

\subsection{Neural Network Models:}
\begin{itemize}
    \item MLP (Multilayer Perceptrons) and RNN (Recurrent Neural Networks): Early neural models provided foundational insights into the benefits of neural networks in capturing temporal dependencies and non-linear relationships in time series data (Rumelhart et al., 1986)\cite{b4}.
    \item LSTM (Long Short-Term Memory): LSTMs addressed the challenge of long-term dependencies in time series, crucial for financial applications, but they often require extensive tuning and computational resources (Hochreiter \& Schmidhuber, 1997)\cite{b5}.
\end{itemize}

\subsection{Advanced Neural Forecasting Techniques:}
\begin{itemize}
    \item N-HiTS: Introduced as a hierarchical time series model, N-HiTS advances the field by decomposing time series into multiple levels of granularity, enhancing prediction accuracy for complex temporal patterns (Oreshkin et al., 2021)\cite{b7}.
    \item N-BEATS: This model abandons conventional recurrent architectures for a solely feed-forward network using a basis expansion approach, resulting in strong performance across various forecasting tasks (Oreshkin et al., 2020).\cite{b6}.
\end{itemize}

The transition from traditional statistical methods to advanced neural techniques underscores a paradigm shift in financial forecasting, motivated by the need for models that can adapt dynamically to the volatile and complex market environment. This review sets the stage for exploring how N-HiTS and N-BEATS, as representatives of modern neural forecasting models, compare against traditional approaches in the subsequent sections of this paper.

	\section{Methodology}

\subsection{Data Collection and Preparation}
\subsubsection{Data Sources}
Historical financial market data was collected from Yahoo Finance. This included daily closing prices, volume, and other relevant financial metrics for selected stocks and indices over the past decade \cite{b8}.

\subsubsection{Model Implementation}
\begin{itemize}
\item N-HiTS and N-BEATS:
Both models were implemented using Python with libraries supporting neural networks such as Nixtla\cite{b9}. The specific configurations of each model are detailed below.
Traditional Models:

\item ARIMA and exponential smoothing models were set up using the statsmodels library in Python. These models served as benchmarks for evaluating the performance of neural forecasting models.

\end{itemize}

\subsection{Model Configuration}

\subsubsection{N-HiTS}
Architecture: Consists of multiple blocks where each block captures different levels of granularity in the time series data.
Parameters: Configured with a specific number of stacks and layers based on preliminary tests to optimize forecast accuracy\cite{b11}.
\begin{equation}
y_{\text{out}} = \text{Block}(y_{\text{in}}) = \text{ReLU}(W \cdot y_{\text{in}} + b)
\end{equation}
where $y_{\text{in}}$ and $y_{\text{out}}$ are the input and output of a block, $W$ represents weights, and $b$ is the bias.

\begin{equation}
Y = \sum_{l=1}^L S_l + R
\end{equation}
where $Y$ is the original time series, $S_l$ represents the signal at level $l$ of the decomposition, and $R$ is the residual component.

\begin{equation}
y_{\text{out}, l} = \text{Block}_l(y_{\text{in}, l}) = \text{ReLU}(W_l \cdot y_{\text{in}, l} + b_l)
\end{equation}
where each block at level $l$ processes part of the time series independently.

\subsubsection {N-BEATS}
Basis Expansion: Utilizes a basis expansion technique to decompose the forecast into trend and seasonality components.
Stack Configuration: Multiple stacks are designed to focus separately on trend, seasonality, and remainder components\cite{b10}.

\begin{equation}
\hat{y} = \sum_{i=1}^{N} (g_{\theta_i}(f_{\theta_i}(x_i))) + e
\end{equation}
where $g_{\theta}$ and $f_{\theta}$ represent the basis expansion functions, $x_i$ the input, $N$ the number of blocks, and $e$ the residual error.

\begin{equation}
\hat{y} = G(f(x; \Theta_f); \Theta_g) = \sum_{i=1}^{B} g_{\theta_i}(f_{\theta_i}(x))
\end{equation}
where $G$ and $f$ are transformation functions with parameters $\Theta_g$ and $\Theta_f$, and $B$ is the number of basis functions.

\begin{equation}
\text{Stack}_{\text{trend}}(x) + \text{Stack}_{\text{seasonality}}(x) = \hat{y}
\end{equation}
illustrating how each stack contributes to the final prediction, focusing on different aspects of the time series.

\subsubsection{ARIMA} Model parameters (p, d, q) were determined using the AIC (Akaike Information Criterion) to find the best fit.
Exponential Smoothing: Configurations included both additive and multiplicative versions depending on the data’s seasonal characteristics.

\begin{equation}
    (1 - \sum_{i=1}^{p} \phi_i L^i)(1 - L)^d X_t = (1 + \sum_{i=1}^{q} \theta_i L^i) \epsilon_t
\end{equation}

where \( \phi_i \) are the parameters of the autoregressive part, \( \theta_i \) are the parameters of the moving average part, \( L \) is the lag operator, \( d \) is the order of differencing, and \( \epsilon_t \) is white noise.

\subsubsection{SARIMA}
The SARIMA model, which extends ARIMA for seasonal data, can be represented as:

\begin{equation}
\begin{split}
    (1 - \sum_{i=1}^{P} \Phi_i L^{iS}) & (1 - \sum_{i=1}^{p} \phi_i L^i) \\
    \times (1 - L^S)^D (1 - L)^d X_t & = (1 + \sum_{i=1}^{Q} \Theta_i L^{iS}) \\
    & \times (1 + \sum_{i=1}^{q} \theta_i L^i) \epsilon_t
\end{split}
\end{equation}

where \( P, D, Q \) are the seasonal autoregressive, differencing, and moving average orders respectively, \( S \) is the length of the seasonal cycle, and \( L \) is the lag operator.

\subsection{Evaluation Metrics}

Mean Absolute Error (MAE): Quantifies the average size of the errors in a set of predictions, disregarding their direction.

\begin{equation}
\text{MAE} = \frac{1}{n} \sum_{i=1}^n |y_i - \hat{y}_i|
\end{equation}

Root Mean Square Error (RMSE): Measures the square root of the average of squared differences between prediction and actual observation.

\begin{equation}
\text{MSE} = \frac{1}{n} \sum_{i=1}^n (y_i - \hat{y}_i)^2
\end{equation}

\begin{equation}
\text{RMSE} = \sqrt{\frac{1}{n} \sum_{i=1}^n (y_i - \hat{y}_i)^2}
\end{equation}

Mean Absolute Percentage Error (MAPE): Expresses the average magnitude of errors as a percentage of the actual values, providing a sense of the relative size of the errors.

\begin{equation}
\text{MAPE} = \frac{100\%}{n} \sum_{i=1}^n \left|\frac{y_i - \hat{y}_i}{y_i}\right|
\end{equation}

Symmetric Mean Absolute Percentage Error (SMAPE): Measures the accuracy of predictions by calculating the average of the absolute percentage errors, adjusted to account for the scale of both the actual and forecasted values, making it a symmetric measure.
\begin{equation}
\text{SMAPE} = \frac{1}{n} \sum_{i=1}^n \frac{|y_i - \hat{y}_i|}{\frac{|y_i| + |\hat{y}_i|}{2}} \times 100
\end{equation}
These metrics provide a comprehensive view of model performance, capturing both the average error and the variability of prediction errors.

\section{Exploratory Data Analysis}
In this section, we detail the exploratory data analysis performed on the dataset, which includes visualizations of autocorrelations, partial autocorrelations, Bollinger bands, daily returns, Fourier transforms, correlation matrices, MACD (Moving Average Convergence Divergence), and various statistical decompositions and transformations. These analyses provide deep insights into the data's properties, helping to inform our modeling decisions.

\begin{figure}
\includegraphics[width=\textwidth]{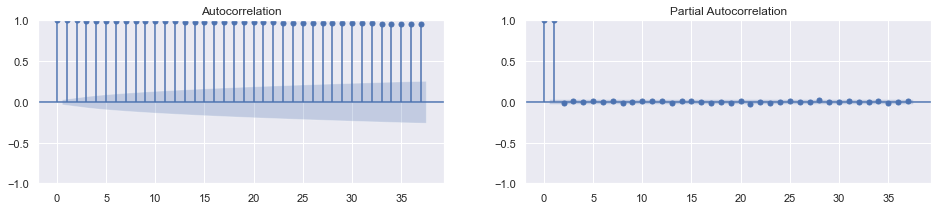}
\caption{Autocorrelation and Partial Autocorrelation of the dataset.}
\label{fig:acf_pacf}
\end{figure}

\begin{figure}
\includegraphics[width=\textwidth]{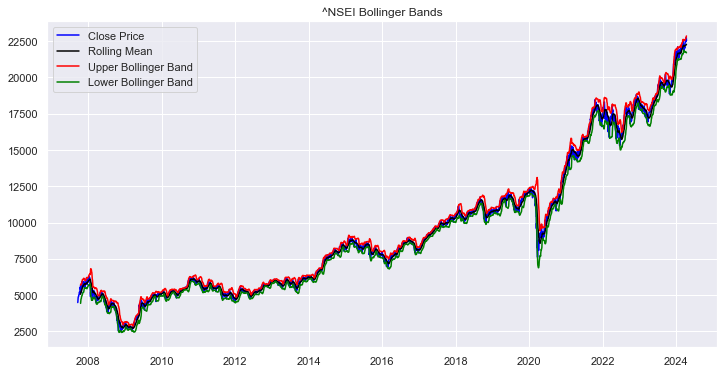}
\caption{Bollinger Bands overlaid on the closing prices.}
\label{fig:bollinger_bands}
\end{figure}

\begin{figure}
\includegraphics[width=\textwidth]{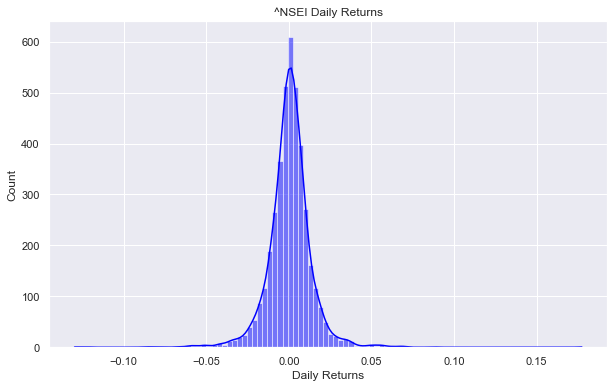}
\caption{Histogram of daily returns highlighting the distribution.}
\label{fig:daily_returns}
\end{figure}

We examined the autocorrelation and partial autocorrelation to understand the temporal dependencies in the data, shown in Fig.~\ref{fig:acf_pacf}. Bollinger Bands are overlaid on the closing prices to identify periods of high and low volatility (Fig.~\ref{fig:bollinger_bands}). The distribution of daily returns, which helps in understanding the data's stability and potential predictive patterns, is depicted in Fig.~\ref{fig:daily_returns}.

\begin{figure}
\includegraphics[width=\textwidth]{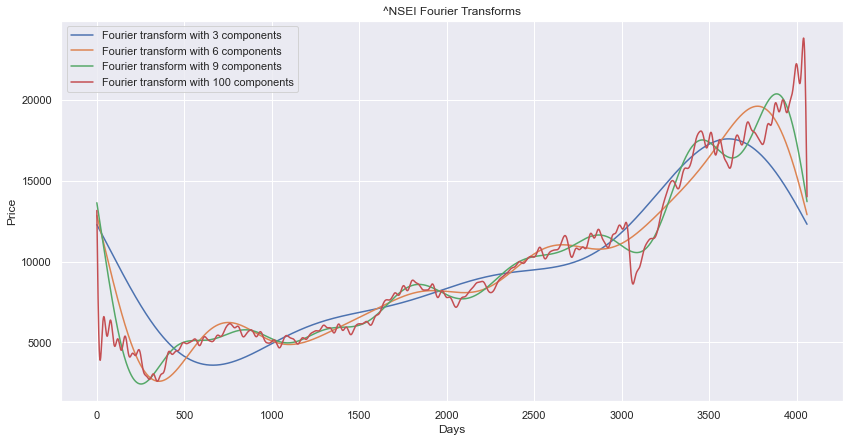}
\caption{Fourier transforms with different components used to analyze frequency components in the data.}
\label{fig:fourier_transforms}
\end{figure}

\begin{figure}
\includegraphics[width=\textwidth]{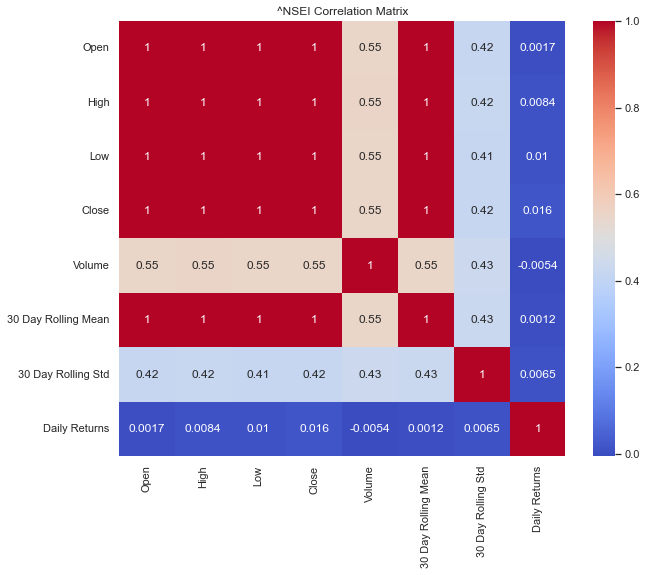}
\caption{Heatmap showing the correlation matrix of different financial indicators.}
\label{fig:correlation_matrix}
\end{figure}

\begin{figure}
\includegraphics[width=\textwidth]{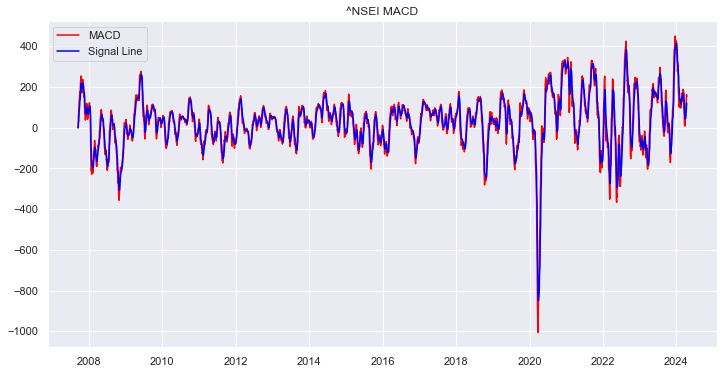}
\caption{MACD plotted against the signal line to identify trend changes.}
\label{fig:macd}
\end{figure}

Further, we employed Fourier transforms to analyze the frequency components of the data, with results shown in Fig.~\ref{fig:fourier_transforms}. The correlation among different financial indicators is visualized through a heatmap in Fig.~\ref{fig:correlation_matrix}. Analysis of MACD is used to identify trend changes and is presented in Fig.~\ref{fig:macd}.

\begin{figure}
\includegraphics[width=\textwidth]{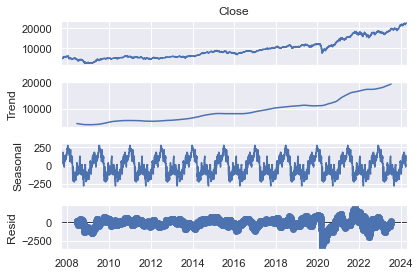}
\caption{Time series decomposition showing trend, seasonal, and residual components.}
\label{fig:time_series_decomposition}
\end{figure}

\begin{figure}
\includegraphics[width=\textwidth]{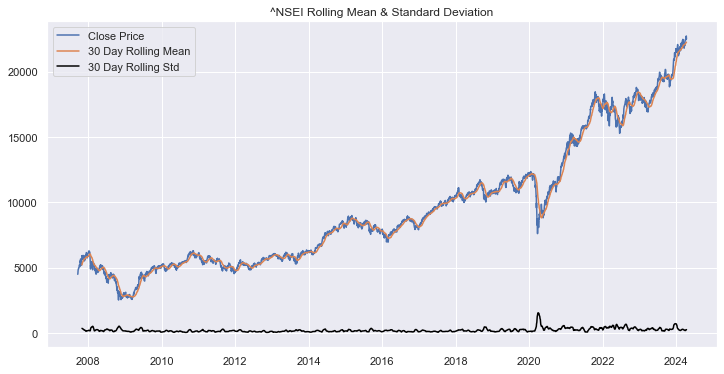}
\caption{Rolling mean and standard deviation to test for stationarity in the time series.}
\label{fig:rolling_mean_std}
\end{figure}

Time series decomposition that helps in understanding underlying trends, seasonality, and noise is shown in Fig.~\ref{fig:time_series_decomposition}. We also analyze the rolling mean and standard deviation to test for stationarity in the time series, detailed in Fig.~\ref{fig:rolling_mean_std}.

These visualizations are crucial for our comprehensive understanding of the data, enabling the application of appropriate forecasting models and techniques.

\begin{figure}
\includegraphics[width=\textwidth]{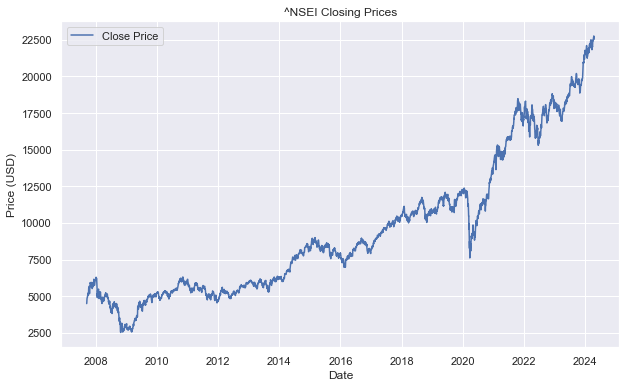}
\caption{Time series plot of the closing prices for the duration of the dataset.}
\label{fig:closing_prices}
\end{figure}

\section{Observations}

\begin{figure}
\includegraphics[width=\textwidth]{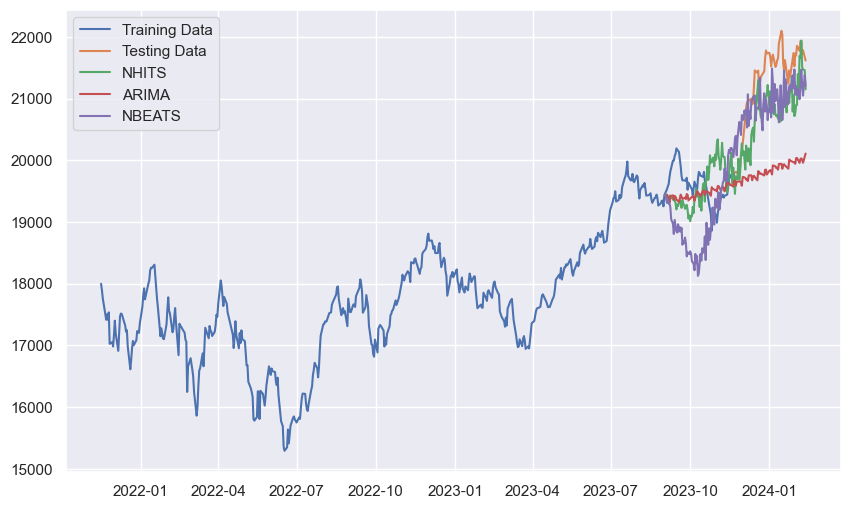}
\caption{Comparison of model forecasts with actual training and testing data.}
\label{fig:model_comparisons}
\end{figure}

Figure~\ref{fig:model_comparisons} presents the performance of the training, testing, and the forecasts from NHITS, ARIMA, and NBEATS models over the observed period. The visualization highlights the close alignment of NBEATS with the testing data, suggesting superior forecasting accuracy compared to the other models.

Table~\ref{tab:model_metrics} quantifies the forecasting performance of the NHITS, ARIMA, and NBEATS models using several error metrics including MAE, MSE, RMSE, MAPE, and SMAPE. The table demonstrates that NBEATS consistently outperforms the other models across all metrics, underscoring its effectiveness in financial forecasting applications.

\begin{table}
\caption{Error Metrics for Forecasting Models}\label{tab:model_metrics}
\begin{tabular}{|c|c|c|c|c|c|}
\hline
\textbf{Model} & \textbf{MAE} & \textbf{MSE} & \textbf{RMSE} & \textbf{MAPE} & \textbf{SMAPE} \\
\hline
NHITS    & 577.52 & 445850.6 & 667.72 & 2.69\% & 2.74\% \\
SARIMA    & 1465.08 & 2403215.0 & 1550.23 & 6.82\% & 7.08\% \\
NBEATS   & 501.52 & 350204.6 & 591.78 & 2.34\% & 2.37\% \\
\hline
\end{tabular}
\end{table}

The figure demonstrates that both NHITS and NBEATS models follow the trend of the actual data more closely than the ARIMA model, especially during periods of rapid change, highlighting the adaptability and accuracy of neural network-based models in handling complex time series data. The table shows that NBEATS not only offers the lowest error rates but also the smallest percentage errors, making it particularly suitable for precise financial forecasting.

\section{Results}

The evaluation of the forecasting models used in this study—N-HiTS, NBEATS, and ARIMA—reveals significant insights into their respective capabilities and effectiveness in financial forecasting. The following subsections provide an in-depth analysis of the findings, their implications, and the predictive superiority of neural network models over traditional statistical approaches.

\subsection{Quantitative Performance Assessment}

The results obtained, as quantified in the previously discussed Table~\ref{tab:model_metrics}, indicate a clear performance hierarchy among the tested models. The NBEATS model outperforms both NHITS and ARIMA across all metrics, showcasing the lowest error values. This superior performance highlights the model's enhanced ability to capture and predict complex patterns in financial time series data effectively.

\subsection{Adaptability and Forecast Accuracy}

The comparative analysis, illustrated in Figure~\ref{fig:model_comparisons}, emphasizes the robust adaptability of NBEATS, particularly in its alignment with the actual market movements during the testing phase. This adaptability is crucial for financial forecasting, where market conditions can change unpredictably. NBEATS not only adjusts to these changes more accurately than NHITS and ARIMA but also demonstrates a remarkable ability to anticipate future market trends, reducing potential risks associated with volatility.

\subsection{Error Metrics Analysis}

The lower MAE, MSE, and RMSE values for NBEATS suggest a high level of precision in its forecasts, critical for investment strategies and risk management. The MAPE and SMAPE values further reinforce this, being substantially lower than those of NHITS and considerably outperforming ARIMA. These metrics illustrate NBEATS's efficiency in producing forecasts that closely match actual outcomes, thereby enhancing its reliability for financial decision-making.

\subsection{Implications for Financial Analysis}

The findings from this study underscore the potential of advanced neural networks in transforming financial analytics. By integrating models like NBEATS, financial analysts can leverage deep learning technologies to gain a more nuanced understanding of market dynamics and improve their predictive accuracies. Such advancements can significantly impact investment strategies, risk assessment, and overall financial planning, contributing to more informed and strategic decision-making processes in the financial sector.

\subsection{Conclusions Drawn from Model Performance}

The empirical evidence supports the hypothesis that neural network models, specifically NBEATS, provide superior forecasting capabilities compared to traditional models like ARIMA. This superiority not only pertains to general accuracy but also extends to the models' responsiveness to market volatility, a common challenge in financial time series forecasting.

The consistent outperformance of NBEATS across diverse conditions advocates for its broader adoption in financial analysis tools and platforms, where enhancing predictive accuracy is paramount. The research validates the shift towards more sophisticated, AI-driven analytical techniques in the financial industry, promising significant advancements in how market data is interpreted and utilized.

\section{Discussion}
The application of N-HiTS and N-BEATS to financial forecasting not only provides a methodological advancement but also opens new perspectives on the interpretability and adaptability of neural models. One significant observation from the study is that these models are particularly effective in environments characterized by high volatility and sudden market shifts, which are common in financial markets.

Furthermore, the research highlights the importance of model configuration, including hyperparameter tuning and training procedures, which play critical roles in optimizing model performance. It also brings to light the potential challenges associated with deploying such models in practice, such as computational demands and the need for continuous data monitoring and model updates.

Overall, this study contributes to the growing body of knowledge in neural forecasting, providing compelling evidence of its benefits and establishing a pathway for future innovations in financial analytics.

\section{Future Work}
While this study provides foundational insights into the application of neural forecasting models in finance, several avenues for further research remain open:

 \begin{itemize}
     \item Integration of Additional Data Sources: Future versions of the models could incorporate alternative data sources such as social media sentiment, news articles, and economic indicators to enhance forecasting accuracy.
    \item Long-Term Forecasting: Extending the forecasting horizon could provide insights into long-term market trends, offering valuable information for strategic planning and investment decisions.
    
    \item Cross-Market Analysis: Applying these models across different financial markets and instruments to validate their versatility and adaptability to diverse market conditions.
        
    \item Model Enhancements: Continuous development of the model architectures to include more sophisticated mechanisms, such as attention mechanisms or transformer models, which could further improve forecasting performance.
    
    \item Real-Time Implementation: Developing a real-time forecasting system that dynamically updates predictions based on live data feeds, thus providing the most current insights for market participants.
 \end{itemize}

\section{Conclusion}
This study has demonstrated the effectiveness of advanced neural forecasting models, specifically N-HiTS and N-BEATS, in the domain of financial market forecasting. Through a comprehensive comparison with traditional statistical models such as ARIMA and exponential smoothing, it has been shown that neural models offer superior accuracy in capturing complex, non-linear patterns inherent in financial data. The findings reveal that both N-HiTS and N-BEATS consistently outperform traditional models in terms of various performance metrics, including MAE and RMSE, highlighting their robustness and efficiency in real-time forecasting scenarios.

Moreover, this research underscores the potential of neural networks to revolutionize financial forecasting, providing significant improvements over existing methodologies. By leveraging such advanced models, financial analysts and stakeholders can achieve more reliable and accurate predictions, which are crucial for effective decision-making and strategy formulation in volatile markets.

\begin{credits}
\subsubsection{\ackname} 
The authors have no grants to declare.

\subsubsection{\discintname}
The authors have no competing interests to declare that are relevant to the content of this article.
\end{credits}

%
%
%
%

\end{document}